\documentclass[fleqn,usenatbib]{mnras}
\usepackage{amsmath}
\usepackage{xcolor, graphicx}
\usepackage{amssymb}
\usepackage{soul}
\usepackage{txfonts}
\usepackage{caption}

\begin{document}

\title[Stellar masses and sizes]{The SLUGGS Survey: stellar masses and effective radii of early-type galaxies from 
Spitzer Space Telescope 3.6$\mu$m imaging} 
\author[D. A. Forbes et al.]{Duncan A. Forbes$^{1}$\thanks{E-mail:
dforbes@swin.edu.au}, Luciana Sinpetru$^{2}$, Giulia Savorgnan$^1$,
Aaron J. Romanowsky$^{3,4}$, 
\newauthor 
Christopher Usher$^{5}$ and Jean Brodie$^{4}$ 
\\
$^{1}$Centre for Astrophysics \& Supercomputing, Swinburne
University, Hawthorn VIC 3122, Australia\\
$^{2}$Institute for Astronomy, University of Edinburgh, Royal Observatory, 
Blackford Hill, Edinburgh EH9 3HJ, UK\\
$^{3}$Department of Physics and Astronomy, San Jos\'e State
University, One Washington Square, San Jose, CA 95192, USA\\
$^{4}$University of California Observatories, 1156 High Street, Santa Cruz, CA 95064, USA\\
$^{5}$Astrophysics Research Institute, Liverpool John Moores University, 146 Brownlow Hill, 
Liverpool L3 5RF, UK\\
}


\pagerange{\pageref{firstpage}--\pageref{lastpage}} \pubyear{2002}

\maketitle

\label{firstpage}

\begin{abstract}

Galaxy starlight at 3.6$\mu$m is an excellent tracer of stellar mass. Here we use the latest 
3.6$\mu$m imaging from the Spitzer Space Telescope 
to measure the total stellar mass and effective radii in a homogeneous way for a sample of galaxies from the 
SLUGGS survey. These galaxies are representative of nearby early-type galaxies in the stellar mass range 
of 10 $<$ log M$_{\ast}$/M$_{\odot}$  $<$ 11.7, and our methodology can be applied to other samples of early-type galaxies. 
We model each galaxy in 2D and estimate its total asymptotic magnitude from a 1D curve-of-growth. Magnitudes are converted into stellar masses using 
a 3.6$\mu$m 
mass-to-light ratio from the latest stellar population models of R\"ock et al., assuming a Kroupa IMF. We apply a ratio based on each galaxy's mean 
mass-weighted stellar age within one effective radius (the mass-to-light ratio is insensitive to galaxy metallicity for the generally old stellar ages 
and high metallicities found in massive early-type galaxies). Our 3.6$\mu$m stellar masses agree well with masses derived from 2.2$\mu$m data.
From the 1D 
surface brightness profile we fit a single Sersic law, excluding the very central regions.  We measure the effective radius, Sersic n parameter and effective surface brightness for each galaxy. We find that galaxy sizes derived from shallow optical imaging and the 2MASS survey 
tend to underestimate the true size of the largest, most massive galaxies in our sample. We adopt the 3.6$\mu$m stellar masses and effective radii for the SLUGGS survey galaxies.

\end{abstract}


\section{Introduction}

The total stellar mass is a fundamental parameter for any galaxy. Not
only do many other galaxy properties vary with stellar mass, but an
accurate measure of stellar mass is required to probe the dark matter content 
(i.e. the total mass minus the stellar mass) in a
galaxy. However, measuring the total stellar mass is problematic, even
once the total luminosity has been accurately measured. For example, a
common approach is to measure the total luminosity of a galaxy at
near-IR wavelengths for which the light mostly comes from old stars
that dominate the mass, and the effects of dust are much reduced compared to
optical wavelengths.  
A typical approach is to use the full-sky ground-based near-IR imaging of the 2MASS 
survey (Jarrett et al. 2003). However it has been reported that the 2MASS reduction pipeline 
systematically underestimates the total luminosity and size of large, nearby galaxies due to 
a truncation of their surface brightness profiles 
(Schombert \& Smith 2012; Scott et al. 2013).

An alternative approach is to use the 3.6$\mu$m
band of the Spitzer Space Telescope (Werner et al. 2004) or the 3.4$\mu$m band of the WISE space telescope 
(Wright et al. 2010).  Such wavelengths are particularly
well suited to measure the stellar masses of galaxies. 
For example, Norris et  al. (2014) concluded that photometry from WISE can {\it ``...provide extremely simple, yet 
robust  stellar mass tracers for dust free older stellar populations..."}. 
This is because the 3.4--3.6$\mu$m light
from galaxies is dominated by the light from old stars, and it is less effected by variations in the 
star formation history than shorter wavelengths.
Although
intermediate-aged stars, hot dust and polycyclic aromatic hydrocarbons
may contribute to the emission at 3.6$\mu$m, these sources are
negligible for most early-type galaxies which are dominated by old stellar populations 
(Meidt et al. 2012; Querejeta et al. 2015).

Here we use 3.6$\mu$m imaging from the Spitzer Space Telescope. 
The 3.6$\mu$m mass-to-light ratio (M/L$_{3.6}$) 
has virtually no dependence on metallicity, and only a 
very small dependence on age for old stellar ages. We use the latest single-burst 
stellar population models (R\"ock et al. 2015) which are based on empirical mid-infrared stellar spectra (Cushing et al. 2005; Rayner et al. 2009).  
These models 
cover a range of metallicity, 
ages and IMF slopes. They are shown to reproduce well the mid-infrared colours of early-type galaxies. 
For a Kroupa IMF, these models give 
M/L$_{3.6}$ $\sim$ 0.8 for a stellar population mean age of 9 Gyr, with a variation between 
different isochrones, i.e. from BaSTI and Padova, of $\sim$0.05. 
For a metallicity range of [Fe/H] = --0.4 to solar (i.e. typical of the mean values for massive early-type galaxies), the variation  
is insignificant at $\sim$0.02.  
We note that the Flexible Stellar Population Synthesis (FSPS; Conroy \& Gunn 2010) models with 
AGB circumstellar dust included (Villaume et al. 2015) also give 
M/L$_{3.6}$ $\sim$0.8 for a 9 Gyr old, moderately metal-rich population.
Meidt et al. (2014) adopted 
a constant value of M/L$_{3.6}$ = 0.6 for their S4G sample, although their sample was dominated by  
late-type galaxies with younger mean ages on average.

Stellar mass-to-light ratios have a strong dependence on the
Initial Mass Function (IMF). The M/L$_{3.6}$ values quoted above refer
to a Kropua IMF. Salpeter and other IMFs tend to have higher
M/L$_{3.6}$ values by a factor of $\sim$1.5-3 (R\"ock et al. 2015), which
would lead to larger stellar masses for a given 3.6$\mu$m
luminosity. Recent work indicates that the IMF for early-type galaxies
is skewed to low mass stars (see e.g. Ferre-Mateu et
al. 2013; Martin-Navarro et al. 2015; McConnell et al. 2016). 
Currently, it is not yet clear what is causing the IMF variations nor whether these variations are confined to 
galaxy central regions, high metallicity regions or spheroids. Here we adopt a 
Kroupa IMF for our global M/L$_{3.6}$ but caution that the stellar
masses for massive elliptical galaxies may need revising upwards. 

Since M/L$_{3.6}$ varies with stellar age, an age-appropriate ratio should be employed. 
Here we adopt an age dependent mass-to-light ratio from the R\"ock et al. (2015) models using 
mean stellar ages from the literature. We assume a Kroupa IMF. 

The SLUGGS survey targets 25 nearby massive early-type galaxies in
different environments and 3 so-called bonus galaxies (Brodie et al. 2014). We study the kinematics
and metallicity of both the galaxy itself, and its system of globular
clusters, to large galactocentric radii.
The sample galaxies are chosen
to cover a range of key parameters including stellar mass and physical size. 
Until now, the approach in the SLUGGS survey to measure stellar mass has
been to obtain the extinction-corrected K-band (2.2$\mu$m) magnitude
from the 2MASS extended galaxy catalog and apply the correction of
Scott et al. (2013) for missing light. We then 
applied a constant 
M/L$_{2.2}$ = 1 irrespective of stellar metallicity or age. A value of unity is simplistic, but 
a reasonable approximation for a very old stellar population 
with a Kroupa IMF (Bruzual \& Charlot 2003). 

The effective radii (R$_e$) of the SLUGGS galaxies are listed in
Brodie et al. (2014, B14), which are based on Cappellari et
al. (2011). Cappellari et al.  used sizes from both optical and
near-IR imaging.  They noted that the near-IR sizes from the 2MASS survey (Jarrett et al. 2003) for the
largest, most massive galaxies appear to be systematically underestimated and 
they scaled-up their near-IR sizes to match the optical sizes on average. A recent study by van den Bosch (2016) 
also found the 2MASS survey to underestimate the sizes (and total fluxes) of nearby galaxies.  Accurate galaxy
effective radii are important in order to compare galaxies on a
similar relative scale.  For example the SLUGGS survey and other
integral field spectroscopy studies, derive kinematic profiles as a
function of effective radii and measure properties such as specific
angular momentum within 1 R$_e$ (Arnold et al. 2014; Alabi et
al. 2015; Foster et al. 2016). Thus the photometric effective radii
need to be accurate in order to correctly compare kinematic properties
between different galaxies. Effective radii, combined with accurate
stellar masses, are needed to probe the dark matter fraction within a
given multiple of R$_e$. The Spitzer Space Telescope imaging presented here offers an opportunity to
revisit the sizes and masses of the SLUGGS early-type galaxies.

In the next sections we present the 3.6$\mu$m data from the Spitzer
Space Telescope and our methodology for deriving total magnitudes,
stellar masses, and effective radii for 27 SLUGGS early-type
galaxies. These new measurements are compared with literature
values. In an Appendix we list the measurements for six additional
nearby early-type galaxies, which we include for the interested reader.

\section{Spitzer Data}

Here we use images from the IRAC instrument of the Spitzer Space Telescope, which has a pixel scale of 1.22 arcsec over a 
5.2 $\times$ 5.2 arcmin$^{-2}$  field-of-view. 
We have downloaded the latest (July 2016) available 3.6$\mu$m basic
calibrated data frames from the Spitzer Heritage Archive.
These Astronomical Observation Requests
(AORs) are detailed in Appendix A (for SLUGGS galaxies) and C (for
non-SLUGGS galaxies). These data have been corrected for scattered light,
dark current, flat-fielded and flux calibrated.  The MOPEX package is
used to assemble the long-exposure ($>$1 s) frames into an image
mosaic showing the field of view around each target galaxy. An example
of the final mosaic for NGC 1407 is shown is Fig. 1.

\begin{figure}
        \includegraphics[angle=-90, width=0.5\textwidth]{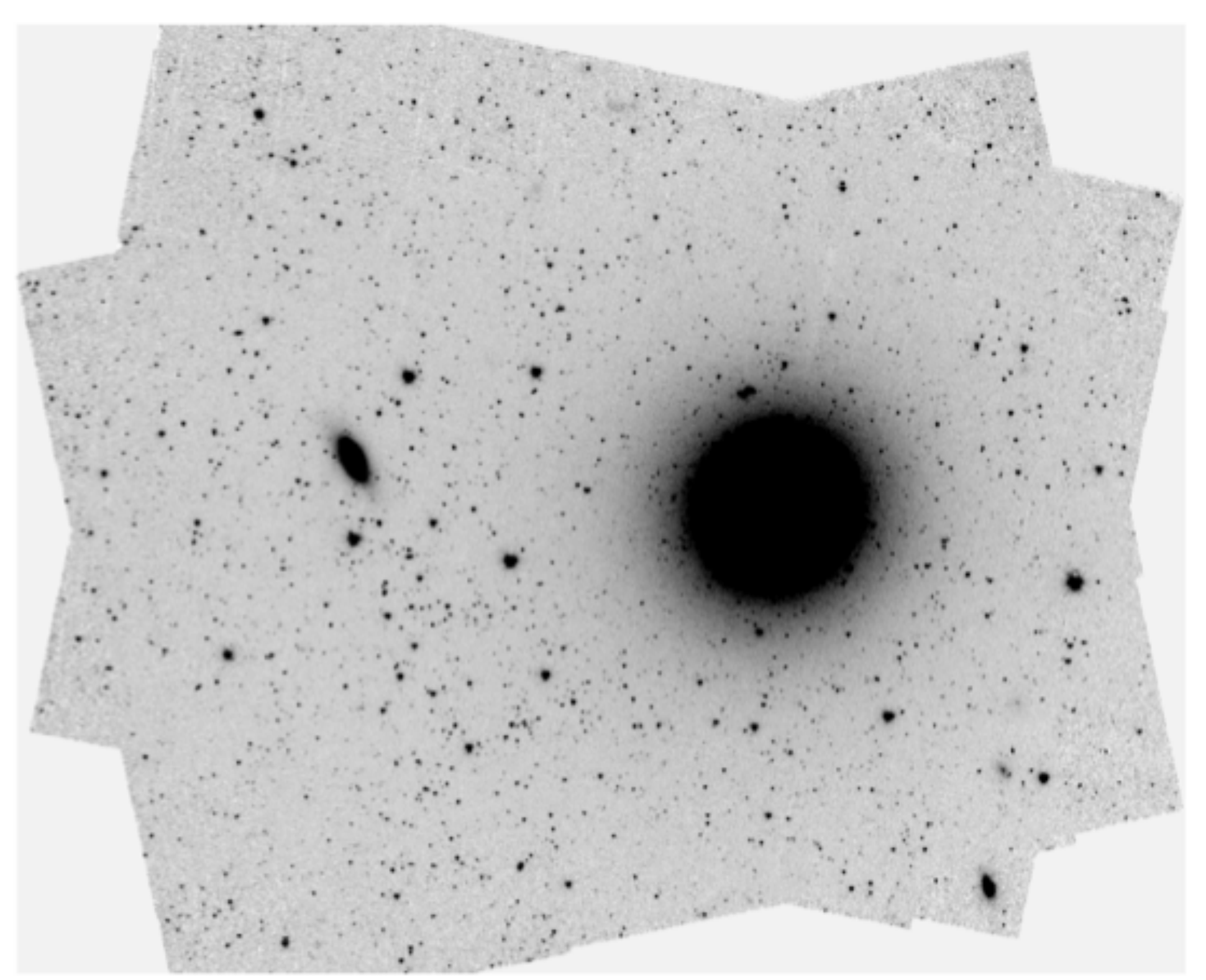}
	    \caption{\label{fig:corrGC2} Spitzer Space Telescope 
3.6$\mu$m image of NGC 1407. North is up and East is left. The galaxy to the North has a projected separation of 
8.5 arcmins from NGC 1407.
}

\end{figure} 

We follow a similar reduction procedure to that of Savorgnan \& Graham (2016, SG16). 
Thus when multiple pointings are available, an overlap correction is applied
to create a uniform background level. Using IRAF 
we determine the sky background level and rms 
at multiple points on the outskirts of each mosaiced
image. The sky values are averaged to give a final value for the sky
background of each image, which is subtracted from the image. 
Finally, bright stars and
other unwanted objects are masked out of the mosaic.  For further details see SG16. 

Unfortunately, the 3.6$\mu$m data for the low-mass SLUGGS galaxy NGC 4474 are
not useful for measuring the total light (and hence mass) or galaxy
size. In this case, the galaxy is only partially visible as it is near the
edge of the available Spitzer pointing.  We therefore adopt its stellar mass (log M$_{\ast}$ = 10.23)  from
its 2MASS K-band magnitude and its effective
radius (R$_e$ = 1.5 kpc) from Brodie et al. (2014). 

\section{Measuring total magnitudes}

To measure the total light from each galaxy we model the galaxy in 2D
using the IRAF task {\it ellipse} and obtain the total magnitude of
the galaxy model.  The galaxy centre was initially allowed to vary,
but if it varied by more than 1 pixel, we fixed it to the average
central value. For a few galaxies (i.e.  NGC 4486, 4594, 5866 and
7457) we could not obtain a good galaxy model with a radially varying
position angle (PA) and so in these cases we fixed the PA to a
representative value based on the radial trend. The model extends in galactocentric radius
until the integrated magnitude at that radius is less than 0.02 mag
different from the previous (penultimate) radius. Thus we effectively
adopt the asymptotic total magnitude of the model galaxy. We estimate
our combined photometric and systematic uncertainty to be $\pm$ 0.05
mag. We do not correct our 3.6$\mu$m magnitudes for Galactic
extinction (which are less than 0.01 mag.).  The total 3.6$\mu$m
magnitudes that we measure in the Vega system, and other basic properties of the 
SLUGGS galaxies, are given in Table 1.

\begin{figure}
        \includegraphics[angle=-90, width=0.5\textwidth]{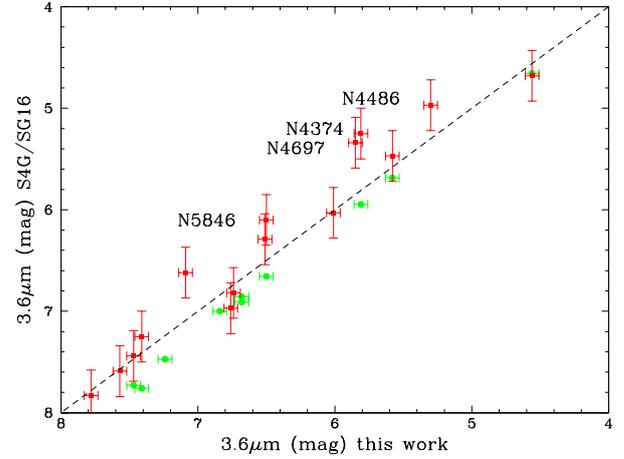}
	    \caption{\label{fig:corrGC2} Comparison of 
3.6$\mu$m magnitude measured in this work with the literature. Data from S4G  are shown by filled green circles and SG16 by filled red squares. 
The dashed line shows the 1:1 line. 
Our 3.6$\mu$m magnitudes lie between those of S4G and SG16. Four SG16 galaxies are highlighted (see text).  
}

\end{figure} 

\begin{table*}
\centering
{\small \caption{SLUGGS galaxy sample and properties}}
\begin{tabular}{@{}cccccccccc}
\hline
\hline
Galaxy & Type & Core & Dist. & Age & m$_{3.6}$ & log M$_{\ast}$ & R$_e$ & $\mu_e$ & n\\
$\rm [NGC]$ & & & [Mpc] & [Gyr] & [mag]  &  [M$_{\odot}$] & [arcsec]  & [mag/arcsec$^2$] & \\
(1) & (2) & (3) & (4) & (5)  & (6) & (7) & (8) & (9) & (10)\\
\hline
 720 & E5 & 1 & 26.9 & 7.8 & 6.92  & 11.27 & 29.1 & 17.54 & 3.8\\
 821 & E6 & 3 & 23.4 & 12.9 & 7.57  & 11.00 & 43.2 & 19.03 & 6.0\\
1023 & S0 & 3 & 11.1 & 13.5 & 6.01 & 10.99 & 48.0 & 17.61 & 4.2\\
1400 & E1/S0 & 1 & 26.8 & 13.8 & 7.44  & 11.08 & 25.6 & 17.87 & 5.0\\
1407 & E0 & 1 & 26.8 & 12.0 & 6.16  & 11.60 & 93.4 & 19.19 & 4.9\\
2768 & E6/S0 & 2 & 21.8 & 13.3 & 6.68  & 11.21 & 60.3 & 18.70 & 3.8\\
2974 & E4/S0 & 3 & 20.9 & 11.8 & 7.47  & 10.93 & 30.2 & 17.99 & 4.3\\
3115 & S0 & 3 & 9.4 & 9.0 & 5.58  & 10.93 & 36.5 & 16.75 & 4.7\\
3377 & E5-6 & 3 & 10.9 & 11.3 & 7.09  & 10.50 & 45.4 & 18.81 & 5.9\\
3607$^\dagger$ & S0 & 1 & 22.2 & 13.5 & 6.51  & 11.39 & 48.2 & 18.33 & 5.3\\
3608 & E1-2 & 1 & 22.3 & 13.0 & 7.41  & 11.03 & 42.9 & 19.00 & 5.3\\
4111 & S0 & -- & 14.6 & 6.0 & 7.24  & 10.52 & 10.1 & 15.71 & 3.0\\
4278 & E1-2 & 1 & 15.6 & 13.7 & 6.84  & 10.95 & 28.3 & 17.51 & 6.2\\
4365 & E3 & 1 & 23.1 & 13.4 & 6.31  & 11.51 & 77.8 & 18.96 & 4.9\\
4374 & E1 & 1 & 18.5 & 13.7 & 5.81  & 11.51 & 139.0 & 19.71 & 8.0\\
4459 & S0 & 3 & 16.0 & 11.9 & 6.76  & 10.98 & 48.3 & 18.52 & 5.4\\
4473 & E5 & 1 & 15.2 & 13.0 & 6.74  & 10.96 & 30.2 & 17.67 & 5.0\\
4474 & S0 & 3 & 15.5 & 11.1 & -- & 10.23 & 17.0 & -- & -- \\
4486 & E0/cD & 1 & 16.7 & 12.7 & 5.30  & 11.62 & 86.6 & 18.24 & 5.1 \\
4494 & E1-2 & 3 & 16.6 & 11.0 & 6.68  & 11.02 & 52.5 & 18.53 & 4.5\\
4526 & S0 & -- & 16.4 & 13.6 & 6.17  & 11.26 & 32.4 & 17.05 & 3.6\\
4564 & E6 & 3 & 15.9 & 13.3 & 7.78  & 10.58 & 14.8 & 16.93 & 3.2\\
4594$^\dagger$ & Sa & 1 & 9.5 & 12.5 & 4.56  & 11.41 & 72.0 & 17.06 & 3.2\\
4649 & E2/S0 & 1 & 16.5 & 13.2 & 5.33  & 11.60 & 79.2 & 18.06 & 4.6\\
4697 & E6 & 3 & 12.5 & 13.4 & 5.85  & 11.15 & 95.8 & 19.08 & 5.3\\
5846 & E0-1/S0 & 1 & 24.2 & 12.7 & 6.50  & 11.46 & 89.8 & 19.38 & 5.2\\
5866$^{\dagger}$ & S0 & -- & 14.9 & 5.9  & 6.50  & 10.83 & 23.4 & 16.59 & 2.8\\
7457 & S0 & 3 & 12.9 & 6.1 & 7.94 & 10.13 & 34.1 & 18.67 & 2.6\\
\hline
\end{tabular}
\begin{flushleft}
{\small 

Notes: columns are (1) galaxy name, $\dagger$ = bonus galaxy, (2) Hubble type, (3)  1= core, 2 = intermediate, 3 = cusp central light profile, (4) distance from B14 (typical uncertainty is $\pm$0.05 dex), (5) mean stellar age from McDermid et al. (2015), see text for exceptions), (6) 3.6 micron apparent magnitude in the Vega system (typical uncertainty is $\pm$0.05), (7) stellar mass (typical uncertainty is $\pm$0.1 dex),  (8) effective radius (typical uncertainty is +0.18 and --0.13  dex), (9) $\mu_e$ (typical uncertainty 
is +0.52 and --1.11 mag.), (10) Sersic n (typical uncertainty is +0.13 and --0.11  dex). Spitzer 3.6$\mu$m imaging is not available for NGC 4474: M$_{\ast}$ and R$_e$ are from 2MASS 2.2$\mu$m imaging.
}

\end{flushleft}
\end{table*}

Total 3.6$\mu$m magnitudes for several SLUGGS galaxies 
from Spitzer data are available in the
S4G study of Munoz-Mateos et al. (2015, S4G) and Savorgnan \& Graham
(2016, SG16). The former study measured total magnitudes using a
curve-of-growth method and adopted the asymptotic magnitude. 
The uncertainty on the S4G total 
magnitudes for the SLUGGS galaxies  is given as $\pm$ 0.001--0.002 mag. in 
the S4G online database. This is likely to be the formal uncertainty on 
fitting their asymptotic magnitude and  does not include other sources of uncertainty.
The latter study fit multiple components to the 1D surface brightness profile of each 
galaxy after 2D modelling. The combination of the different components was used to calculate the 
total magnitude, with an estimated uncertainty of $\pm$0.25 mag. 

Fig. 2 shows a comparison, for the SLUGGS galaxies, of our measured
magnitude against 3.6$\mu$m total magnitudes from S4G (converted to the Vega system) and SG16. 
Our magnitudes generally lie in between these two studies. 
We are
systematically brighter than S4G by $\sim$0.18 mag. on average.
Our measurements agree fairly well with SG16 with the exception of 
four galaxies that are more than 
0.3 mag brighter than us  
(and have even larger discrepancies with S4G magnitudes when in common). 
Three of the galaxies, NGC 5846, NGC 4374, and NGC 4486, feature a partially
depleted core at their centre (Lauer et al. 2007; Krajnovic et al. 2013).   
The total magnitudes of SG16
do not take this into account, i.e they masked out the core and fit 
a Sersic profile (rather than a core-Sersic profile) to the galaxy light. 
This effectively overestimates the total magnitude of each galaxy by the 
amount of light `missing' due to the depleted core, i.e. 
$\sim$0.2--0.3 mag. For NGC 4697 we suspect that the best-fit profiles of SG16 
overestimated the spheroid effective radius and consequently
its luminosity. Their 1D surface brightness profile for this galaxy is
less extended than the best-fit effective radius itself. 


\section{Calculating total stellar masses}

\begin{figure}
        \includegraphics[angle=-90, width=0.5\textwidth]{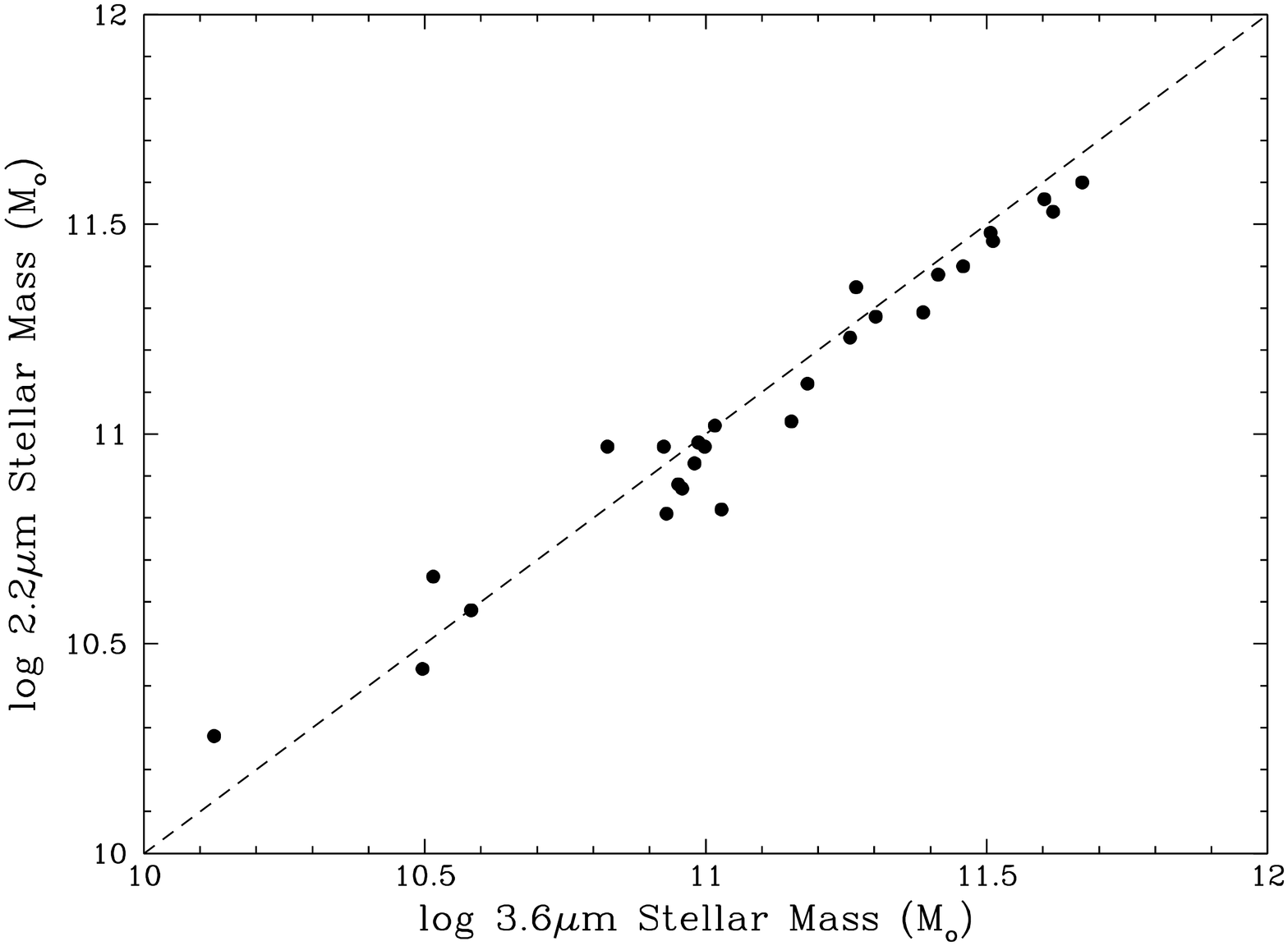}
	    \caption{\label{fig:corrGC2} Comparison of 
3.6$\mu$m stellar mass measured in this work with the K-band (2.2$\mu$m) 
stellar mass. The dashed line shows the 1:1 line.  
}

\end{figure} 

To calculate the total 3.6$\mu$m luminosity in solar units in the Vega system for 
each galaxy we convert the 
3.6$\mu$m apparent magnitude into a total luminosity assuming that the 
absolute magnitude of the Sun to be
M$_{3.6}$ = 3.24 (Oh et al. 2008), 
and taking its distance from Table 1.  We note that the distances, usually based on surface brightness fluctuations, contribute an uncertainty of 
$\sim$0.1 dex to the luminosity. 

We multiply the luminosity by the 3.6$\mu$m mass-to-light ratio from the single stellar population models of R\"ock et al. (2015).
In particular, we use M/L$_{3.6}$  appropriate for each galaxy's mean mass-weighted 
stellar age within 1 R$_e$ taken from McDermid et al. (2015), supplemented by values from Rembold et al. 
(2005) for NGC 720, Norris et al. (2006) for NGC 3115, 
Spolaor et al. (2008) for NGC 1400 and 1407, and Sanchez-Blazquez et al. (2006) for NGC 4594. These ages are given in Table 1. 
The 3.6$\mu$m mass-to-light ratio that we apply varies from $\sim$0.60 to 1.0.
We use the Padova isochrones for solar metallicity (which is a reasonable value for 
our early-type galaxies within 1 R$_e$; McDermid et al. 2015). We note that the equivalent 
BaSTI isochrones differ by only 0.01 for our typical age.
We estimate that the 
uncertainty in M/L$_{3.6}$ due to differences in isochrone tracks ($\pm$0.03), mean age uncertainty ($\pm$0.05) and metallicity variations ($\pm$0.02), combined
with our measurement and distance uncertainties, give 
a final uncertainty of about 0.1 dex in log stellar mass. We also 
assume a Kroupa IMF. As noted in the Introduction, although there is evidence for an  IMF skewed to low mass stars, the effect seems largely 
limited to the central regions of the most massive galaxies. Nevertheless this gives rise 
to a systematic underestimate of the stellar masses of the most massive galaxies.

Each galaxy in this study has a total stellar mass determined from the 
total K-band (2.2$\mu$m) 
magnitude from the 2MASS survey. The 2MASS 2.2$\mu$m magnitude is corrected for missing flux according to Scott et al. (2013) 
and  we take the absolute magnitude of the Sun to be M$_{2.2}$ = 3.28 (table 2.1 from Binney \& Merrifield 1998). 
The stellar mass has been calculated in previous SLUGGS papers  
assuming a fixed M/L$_{2.2}$ = 1.0 irrespective of stellar age (e.g. Alabi et al. 2016). 
A value of unity is reasonably representative of an old, metal-rich stellar 
population with a Kroupa-like IMF. 

In Fig. 3 we show a comparison of our new 3.6$\mu$m-based stellar masses 
versus the stellar mass obtained from the K-band. We find an excellent overall correspondence with the 
stellar masses derived using the K-band. Galaxies with lower 
3.6$\mu$m masses relative to the previous 2.2$\mu$m masses (i.e. that lie above the unity line) tend to be those 
with young ($\sim$6 Gyr) mean stellar ages.  
The overall excellent agreement indicates that both 3.6$\mu$m and 2.2$\mu$m total magnitudes give reliable stellar masses (under the
same assumption of a Kroupa IMF).  Given the small difference in 2.2 vs 3.6 $\mu$m stellar masses, we will continue to adopt 
the K-band stellar mass of log M$_{\ast}$ = 10.23 for NGC 4474 for which we are unable to measure a 3.6$\mu$m magnitude. 

\section{Measuring galaxy sizes}

Each galaxy surface brightness profile is fit with a single Sersic law 
(Graham \& Driver 2005). We exclude the inner 2 pixels (2.44 arcsec), i.e. we only fit radii that are larger 
than the effective FWHM resolution of the Spitzer Space Telescope. This also means that the presence of any 
nuclear star cluster or AGN does not affect the fits. Most of our 
galaxies reveal central surface brightness profiles that can be well described as either a core or a cusp (Lauer et al. 2007; Krajnovic et al. 2013; Dullo \& Graham 2013), as listed in Table 1. In the case of cusps, they are generally well fit by a single Sersic profile. On the other hand for core profiles we exclude the so-called depleted core region from the fits. Thus the fitting range for each galaxy is either $>$2 pixels for the cusp, intermediate and unknown 
galaxies and greater than the depleted core region for the core galaxies. 

The fits to the 3.6$\mu$m surface brightness profile for each SLUGGS galaxy are shown in Appendix B. The central region excluded from each fit is indicated by open circles, and can be most clearly seen in the Sersic profile minus data residual profiles. The code used for the fitting process is the same as SG16. For most galaxies SG16 fit multiple components to each galaxy. However, they did fit a single Sersic profile to three large SLUGGS galaxies. In these cases our effective radii agree very well with their value, i.e. NGC 4374: 139.0 vs 129.8 arcsec, NGC 4486: 86.6 vs 87.1 arcsec and NGC 5846: 89.8 vs 83.4 arcsec.

We list the (equivalent circular) effective radii from the single Sersic fits to our 3.6$\mu$m surface brightness profiles in Table 1. 
A typical uncertainty associated with the effective radius of the SLUGGS galaxies is calculated based on the average of the uncertainties of the 14 early-type galaxies effective radii in the sample of SG16 with single Sersic fits compared to those from other studies (see SG16 for details). The 1$\sigma$ uncertainty of +0.18 dex and --0.13 dex thus takes into account both random and systematic errors. If we only considered random measurement errors a smaller uncertainty would result. 
Table 1 also lists the other fitting parameters, i.e. the Sersic n value and the surface brightness at the effective radius $\mu_e$. Again, we adopt 
the typical uncertainties found for the 14 early-type galaxies, i.e. +0.13 and --0.11 dex on n, and +0.52 and --1.11 mag on $\mu_e$. We note that Sersic parameters are strongly correlated with each other. 

In Fig. 4 we compare our new effective radii from the Spitzer images with those listed by Brodie et al. (2014), i.e. from Cappellari et al. (2011), based on optical data. We find good correspondence for small-sized galaxies and generally we measure greater effective radii for the large-sized galaxies. This 
implies that the sizes of the most massive galaxies are underestimated, as was suggested by Cappellari et al. (2011). Given the reasonable 
agreement in sizes for the smaller galaxies, the uncertainties from the work of SG16 may be an overestimate. 

Fig. 4 also shows the effective radii from the 2MASS Large Galaxy Atlas (Jarrett et al. 2003). In particular, we take the K-band effective radius along the 
semi-major axis ($k\_r\_eff$) and multiply it by an ellipticity correction ($\sqrt(k\_ba)$ to obtain an equivalent circular R$_e$ value.  Cappellari et al. (2011) found that the 2MASS effective radii were on average 1.7$\times$ smaller than the optical radii from the RC3. The dotted line in this plot indicates sizes reduced by this factor and indeed the 2MASS R$_e$ values scatter about this line. 
The smaller sizes may be due to an oversubtraction of the sky background which truncates the 2MASS K-band 
surface brightness profile (Schombert \& Smith 2012). 

\begin{table}
\centering
{\small \caption{Massive Galaxy Sizes}}
\begin{tabular}{@{}ccccccr}
\hline
\hline
Galaxy & 3.6$\mu$m & B14 & Kor09 & Chen10 & Vika13 & Other \\
$\rm [NGC]$ & [arcsec] & [arcsec] & [arcsec] & [arcsec] & [arcsec] & \\\
(1) & (2) & (3) & (4) & (5)  & (6) & (7) \\
\hline
1407 & 93 & 63 & -- & -- & -- & 100 (P15)\\ 
4365 & 78 & 53 & $\sim$154 & 97 & 78 & 126 (B12)\\
4374 & 139 & 53 & $\sim$135 & 131 & 96 & --\\
4486 & 87 & 81 & $\sim$630 & 105 & 82 & --\\
4649 & 79 & 66 & $\sim$118 & 105 & 68 & --\\
5846 & 90 & 59 & -- & -- & -- & 81 (N14)\\
\hline
\end{tabular}
\begin{flushleft}
{\small 

Notes: columns are (1) galaxy name, effective radii in arcsec from (2) 3.6$\mu$m imaging (this work), (3) Brodie et al. (2014), 
(4) Kormendy et al. (2009), (5) Chen et al. (2010), (6) Vika et al. (2013), and (7) other published SLUGGS works (i.e. 
Pota et al. 2015; Blom et al. 2012 and Napolitano et al. 2014).
}

\end{flushleft}
\end{table}

We find that both the 2MASS and B14 effective radii
underestimate the true size of the most massive, largest galaxies.  In
Table 2 we list R$_e$ values in arcsec from single Sersic fits to the
surface brightness profiles of the six most massive SLUGGS galaxies from the literature,
along with the B14 values and those measured in this work. 
We  include the single Sersic fits to SDSS imaging from Chen et al. (2010) and Vika et al. (2013). In the case of Vika et al. 
we quote the z band R$_e$ value. The final column in Table 2 gives R$_e$ values derived by previous SLUGGS studies of 
Blom et al. (2012) and Pota et al. 
(2015) from deep optical imaging, and that adopted by Napolitano et al. (2014).

We also include effective radii from Kormendy et
al. (2009) who presented very deep optical imaging from multiple sources of
Virgo cluster galaxies, and fit single Sersic profiles excluding the
central core region (as we have done). We have converted their
semi-major axis effective radii to equivalent circular ones using each
galaxy's average ellipticity. 
We also have four low mass Virgo galaxies in common with 
Kormendy et al. and we find good agreement for three (NGC 4473, 4564 and 4551), 
For NGC 4459 we find a somewhat larger size than
Kormendy et al. (2009) but this galaxy is difficult to model correctly
given the bright nearby foreground star (despite our efforts to mask
it from the 2D modelling process).

For the massive Virgo galaxies, we have excellent agreement for NGC 4374 (M84)
with Kormendy et al. and Chen et al.  For NGC 4365, 4486 and 4649
(M60), we have good agreement with Vika et al., but we find systematically 
smaller sizes than Chen et al., and significantly smaller sizes than
Kormendy et al. (and Blom et al. 2012 for NGC 4365). We suspect that this is because these galaxies have 
elongated isophotes in their outer regions, i.e light that goes beyond
the Spitzer mosaic (whereas NGC 4374 has very circular outer isophotes
and is well contained within our mosaic).  For NGC 4486 (M87)
Kormendy et al. derive a size that is 6--8$\times$ that of other
studies (corresponding to $\sim$50 kpc). Their single Sersic fit
includes the extended light of the cD envelope.  Our measured
3.6$\mu$m effective radii for NGC 1407 and NGC 5846 are similar to
those used by Pota et al. (2015) and Napolitano et al. (2014) but
significantly larger than B14.

\begin{figure}
        \includegraphics[angle=-90, width=0.5\textwidth]{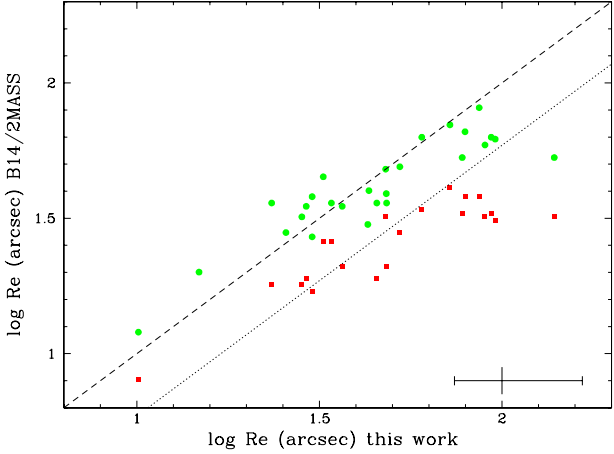}
	    \caption{\label{fig:corrGC2} Comparison of the 
3.6$\mu$m effective radius measured in this work with other works. 
The optical-based sizes used by Brodie et al. (2014, B14) are shown by filled 
green circles, and sizes from the near-IR 2MASS LGA are shown by filled red squares.
The dashed line shows the 1:1 line, whereas the dotted line shows a reduction by a factor of 1.7$\times$ from the unity relation.  
A typical error bar for our effective radii measurements is shown at lower right. The optical and 2MASS near-IR sizes of the large galaxies appear to be 
underestimated. 
}

\end{figure} 


We compare our 3.6$\mu$m sizes and stellar masses with the Virgo cluster galaxies of Chen et al. (2010) in Fig. 5. 
Chen et al. fit single Sersic profiles to multi-filter SDSS imaging of $\sim$ 100 Virgo cluster early-type galaxies. 
We convert their measurements into physical properties assuming a Virgo distance of 16.5 Mpc, and log M/L$_g$ = 0.7 
from Bell et al. (2003) for red galaxies. Fig. 5 shows that the distribution of sizes and stellar masses from 
our 3.6$\mu$m measurements for the SLUGGS galaxies are consistent with that of the 
Virgo early-type galaxies from Chen et al. (2010). Thus we expect our new stellar masses and effective radii for the SLUGGS galaxies to be 
representative of nearby early-type galaxies in general. 

It is clear from Table 2 that a wide variety of galaxy effective radii can be found in the literature for massive galaxies 
(even when restricted to 
a single Sersic fit). 
We recognise that our effective radii derived from Spitzer 3.6$\mu$m imaging may be updated by 
deeper imaging studies (perhaps leading to even larger R$_e$ values if a single Sersic continues to be adopted), however  
our Spitzer imaging provides a homogenous set of improved effective radii for the SLUGGS survey which we now adopt.

\begin{figure}
        \includegraphics[angle=-90, width=0.5\textwidth]{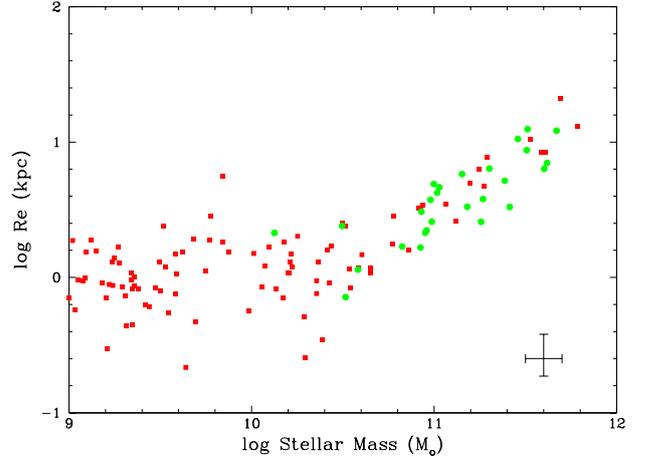}
	    \caption{\label{fig:corrGC2} Size--mass distribution. We show data from Chen et al. (2010) as filled red squares and measurements from this work using the Spitzer Space Telescope imaging as filled green circles. A typical errorbar is shown lower right. Overall the two distributions are very similar. 
}

\end{figure} 

\section{Conclusions}

Using archival Spitzer Space Telescope imaging of the nearby early-type galaxies from the SLUGGS survey, we have created 3.6$\mu$m mosaic images of 
each galaxy (excluding NGC 4474 for which Spitzer imaging was not available). After masking out foreground stars and other unwanted features, we modelled each galaxy and derived total asymptotic magnitudes. Our total magnitudes generally lie in between those of 3.6$\mu$m literature studies for the galaxies in common. 
These total magnitudes are converted into stellar masses using the single stellar population models of R\"ock et al. (2015). The 3.6$\mu$m flux from early-type galaxies is an ideal tracer of stellar mass as the mass-to-light ratio at this wavelength is very insensitive to metallicity and only mildly sensitive to age for old stellar populations. Here we use M/L$_{3.6}$ from Padova isochrones that is adjusted for each galaxy's mean stellar age while assuming a Kroupa IMF. We estimate the final uncertainty  in log stellar mass to be $\pm$0.1 dex.  We find that our 3.6$\mu$m stellar masses have a strong linear correlation with stellar masses derived at 2.2$\mu$m, after correcting the 2MASS K-band fluxes for missing light.

From our 2D galaxy modelling, we fit a single Sersic law to the 3.6$\mu$m surface brightness profile in 1D. We exclude the central 2 pixels in all galaxies 
(corresponding to the FWHM of the Spitzer Space Telescope) and in addition exclude the central core of the massive galaxies that contain 
so-called depleted cores. 
As well as new effective radii (R$_e$), we derive the surface brightness at 1 R$_e$ and the Sersic n parameter. Our 3.6$\mu$m sizes show good agreement with literature values from optical imaging as used by Brodie et al. (2014), and show that sizes from the 2MASS (K-band) survey systematically underestimate the true size (as found previously by 
Cappellari et al. 2011). For the larger, more massive galaxies we find that the optical sizes used by Brodie et al. (2014) 
are also systematically underestimated relative to the sizes from our 3.6$\mu$m imaging and deep optical imaging in the literature. Our new sizes and stellar masses show good agreement with those of Virgo cluster early-type galaxies measured from SDSS imaging. We now adopt the sizes and stellar masses, measured in a homogeneous way from our 3.6$\mu$m imaging, for galaxies in the SLUGGS survey. Our methodology can be adopted by other studies requiring more accurate effective radii and stellar masses for nearby early-type galaxies.

\section{Acknowledgements}

We thank A. Ferre-Mateu and A. Alabi for useful comments. We thank the referee for several
suggestions that have improved the paper. 
DAF thanks the ARC for financial support via DP130100388. JB and AR acknowledge NSF 
grants AST-1211995, AST-1616598 and AST-1616710. 
This work is based on observations made with the Spitzer Space Telescope, 
which is operated by the Jet Propulsion Laboratory, California Institute of 
Technology under a contract with NASA.

\section{References}

Alabi A.~B., et al., 2015, MNRAS, 452, 2208 \\
Arnold J.~A., et al., 2014, ApJ, 791, 80\\
Binney J., Merrifield M., 1998, Galactic Astronomy, Princeton University Press\\ 
Blom C., Spitler L.~R., Forbes D.~A., 2012, MNRAS, 420, 37 \\
Brodie J.~P., et al., 2014, ApJ, 796, 52 (B14)\\
Bruzual G., Charlot S., 2003, MNRAS, 344, 1000\\
Cappellari M., et al., 2011, MNRAS, 413, 813 \\
Chen C.-W., C{\^o}t{\'e} P., West A.~A., Peng E.~W., Ferrarese L., 2010, ApJS, 191, 1 \\
Conroy C., Gunn J.~E., 2010, ApJ, 712, 833\\
Cushing M.~C., Rayner J.~T., Vacca W.~D., 2005, ApJ, 623, 1115 \\
Dullo B.~T., Graham A.~W., 2013, ApJ, 768, 36 \\
Ferr{\'e}-Mateu A., Vazdekis A., de la Rosa I.~G., 2013, MNRAS, 431, 440 \\
Foster C., et al., 2016, MNRAS, 457, 147\\
Graham A.~W., Driver S.~P., 2005, PASA, 22, 118\\
Jarrett T.~H., Chester T., Cutri R., Schneider S.~E., Huchra J.~P., 2003, AJ, 125, 525 \\
Kormendy J., Fisher D.~B., Cornell M.~E., Bender R., 2009, ApJS, 182, 216 \\
Krajnovi{\'c} D., et al., 2013, MNRAS, 433, 2812 \\
Lauer T.~R., et al., 2007, ApJ, 664, 226 \\
Mart{\'{\i}}n-Navarro I., Barbera F.~L., Vazdekis A., Falc{\'o}n-Barroso J., Ferreras I., 2015, MNRAS, 447, 1033 \\
Meidt S E., et al., 2012, ApJ, 744, 17 \\
Meidt S.~E., et al., 2014, ApJ, 788, 144 \\
McConnell N.~J., Lu J.~R., Mann A.~W., 2016, ApJ, 821, 39 \\
McDermid R.~M., et al., 2015, MNRAS, 448, 3484 \\
Milone A.~D.~C., Rickes M.~G., Pastoriza M.~G., 2007, A\&A, 469, 89 \\
Mu{\~n}oz-Mateos J.~C., et al., 2015, ApJS, 219, 3 (S4G)\\
Napolitano N.~R., Pota V., Romanowsky A.~J., Forbes D.~A., Brodie J.~P., Foster C., 2014, MNRAS, 439, 659 \\
Norris M.~A., Sharples R.~M., Kuntschner H., 2006, MNRAS, 367, 815 \\
Norris M.~A., Meidt S., Van de Ven G., Schinnerer E., Groves B., Querejeta M., 2014, ApJ, 797, 55 \\
Oh S.-H., de Blok W.~J.~G., Walter F., Brinks E., Kennicutt R.~C., Jr., 2008, AJ, 136, 2761\\
Pota V., et al., 2015, MNRAS, 450, 3345 \\
Querejeta M., et al., 2015, ApJS, 219, 5\\
Rayner J.~T., Cushing M.~C., Vacca W.~D., 2009, ApJS, 185, 289 \\
Rembold S.~B., Pastoriza M.~G., Bruzual G., 2005, A\&A, 436, 57 \\
R{\"o}ck B., Vazdekis A., Peletier R.~F., Knapen J.~H., Falc{\'o}n-Barroso J., 2015, MNRAS, 449, 2853 \\
Sanchez-Blazquez P., Gorgas J., Cardiel N., Gonzalez J.~J., 2006, A\&A, 457, 809 \\
Savorgnan G.~A.~D., Graham A.~W., 2016, ApJS, 222, 10 (SG16)\\
Schombert J., Smith A.~K., 2012, PASA, 29, 174 \\
Scott N., Graham A.~W., Schombert J., 2013, ApJ, 768, 76 \\
Spolaor M., Forbes D.~A., Proctor R.~N., Hau G.~K.~T., Brough S., 2008, MNRAS, 385, 675 \\
Vika M., Bamford S.~P., H{\"a}u{\ss}ler B., Rojas A.~L., Borch A., Nichol R.~C., 2013, MNRAS, 435, 623 \\
Villaume A., Conroy C., Johnson B.~D., 2015, ApJ, 806, 82 \\
Werner M.~W., et al., 2004, ApJS, 154, 1 \\
Wright E.~L., et al., 2010, AJ, 140, 1868\\
van den Bosch R., 2016, arXiv:1606.01246\\

\section{Appendix A}

Summary of Astronomical Observation Requests (AORs) of SLUGGS galaxies downloaded in July 2016 from the Spitzer Heritage Archive (http://sha.ipac.caltech.edu) are given in Table 3.

\begin{table*}
\centering
{\small \caption{Spitzer Space Telescope Astronomical Observation Requests for SLUGGS Galaxies}}
\begin{tabular}{@{}cl}
\hline
\hline
Galaxy & Astronomical Observation Requests\\
$\rm [NGC]$ & \\
\hline
720 &
	r49345024, 
	r49345280\\
	
821	&
	r14569216, 
	r49418752, 
	r49419008\\
	
1023 &
	r4432640,
	r50631168,
	r50631680,
	r50631936,
	r52778496,
	r52778752,
	r52779008,
	r52779264,
	r52779520,
	r52779776,
	r52780032\\

1400 &
	r49436416\\
	
1407	 &
	r49348096,
	r49348352\\

2768 &
	r18031872\\
	
2974 &
	r18032384,
	r49613056\\
	
3115 &
	r4441088\\
	
3377 &
	r4444928,
	r49411328,
	r50545664,
	r50545920,
	r50546176,
	r52910336,
	r52910592,
	r52910848,
	r52911104,
	r52911360,
	r52911616,
	r52911872\\

3607 &
	r4449536,
	r49389312,
	r49614592,
	r49614848\\

3608 &
	r18033408,
	r49460736,
	r49614848\\
	
4111 &
	r30984192,
	r31015424,
	r42249216,
	r42249472,
	r50528000,
	r50528256,
	r50528512,
	r52912128,
	r52912384,
	r52912640,
	r52912896,
	r52913152,
	r52913408\\

4278 &
	r4461568,
	r49616128\\
	
4365	 &
	r11115264,
	r49358336,
	r49358592,
	r50576640,
	r50577152,
	r50577664,
	r52976640,
	r52976896,
	r52977152,
	r52977408,
	r52977664,
	r52977920\\
	
4374 &
	r4463872,
	r50608128,
	r50608640,
	r50609152,
	r52971264,
	r52971520,
	r52971776,
	r52972032,
	r52972288,
	r52972544\\

4459 &
	r11378944,
	r49501696,
	r49501952\\

4473 &
	r11377920,
	r49339904,
	r49340160,
	r50554368,
	r50554880,
	r50555392,
	r53005312,
	r53005568,
	r53005824,
	r53006080,
	r53006336,
	r53006592\\

4474 &
--\\
	
4486	 &
	r12673792,
	r49337856,
	r49338112,
	r50576384,
	r50576896,
	r50577408,
	r52962304,
	r52962560,
	r52962816,
	r52963072,
	r52963328,
	r52963584\\
	
4494 &
	r18035200\\

4526 &
	r4472064,
	r49341440,
	r49341696,
	r49595904,
	r50644736,
	r50644992,
	r50645248,
	r52992768,
	r52993024,
	r52993280,
	r52993792,
	r52994048\\

4564 &
	r14572032,
	r49510912,
	r49511168,
	r50647040,
	r50647296,
	r50647552,
	r52942592,
	r52942848,
	r52943104,
	r52943360,
	r52943616,
	r52943872\\

4594 &
	r5517824,
	r5518080,
	r50595328,
	r50595840,
	r50596864,
	r52765696,
	r52765952,
	r52766208,
	r52766464,
	r52766720,
	r52766976,
	r52767232\\

4649 &
	r4476672,
	r49337344,
	r49337600,
	r50590208,
	r50590976,
	r50591488,
	r52967680,
	r52967936,
	r52968192,
	r52968448,
	r52968704,
	r52968960\\
	
4697 &
	r10896896,
	r49359872,
	r49360128,
	r50622464,
	r50622720,
	r50622976,
	r52809472,
	r52809728,
	r52809984,
	r52810240,
	r52810496,
	r52810752,
	r52811008\\
	
5846 &
	r4491264,
	r16310272,
	r49363968,
	r49364224\\

5866 &
	r5526016,
	r5526272\\

7457	 &
	r18037504,
	r50547200,
	r50547456,
	r50547712,
	r52946176,
	r52946432,
	r52946688,
	r52946944,
	r52947200,
	r52947456\\
	
\hline
\end{tabular}
\begin{flushleft}
{\small 

}

\end{flushleft}
\end{table*}

\section{Appendix B}

Sersic fits to 3.6$\mu$m surface brightness profiles of SLUGGS galaxies are shown in Figures 6 to 14.

\begin{figure}
        \includegraphics[angle=0, width=0.40\textwidth]{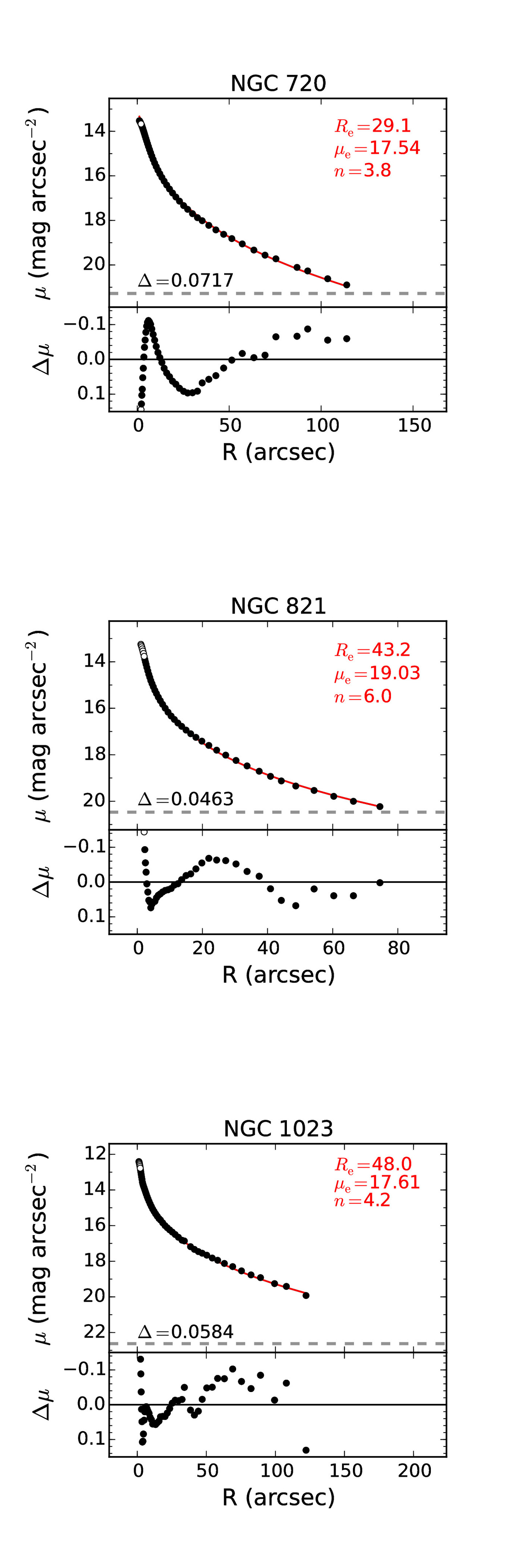}
	    \caption{\label{fig:corrGC2} Sersic fit to 3.6$\mu$m surface brightness profile and residuals as a function of circular equivalent radius. The {\it upper} panel shows the data points (with  
	    excluded data points shown by open circles) and the best fit Sersic profile in red. Parameters for the Sersic fit are given in the top right.  The dashed line 
	    shows 3$\times$ the rms of the sky background level. $\Delta$ gives the rms of the residuals in mag arcsec$^{-2}$. The {\it lower} panel shows the residuals of the Sersic model fit minus the surface brightness data. 
}

\end{figure}

\begin{figure}
        \includegraphics[angle=0, width=0.40\textwidth]{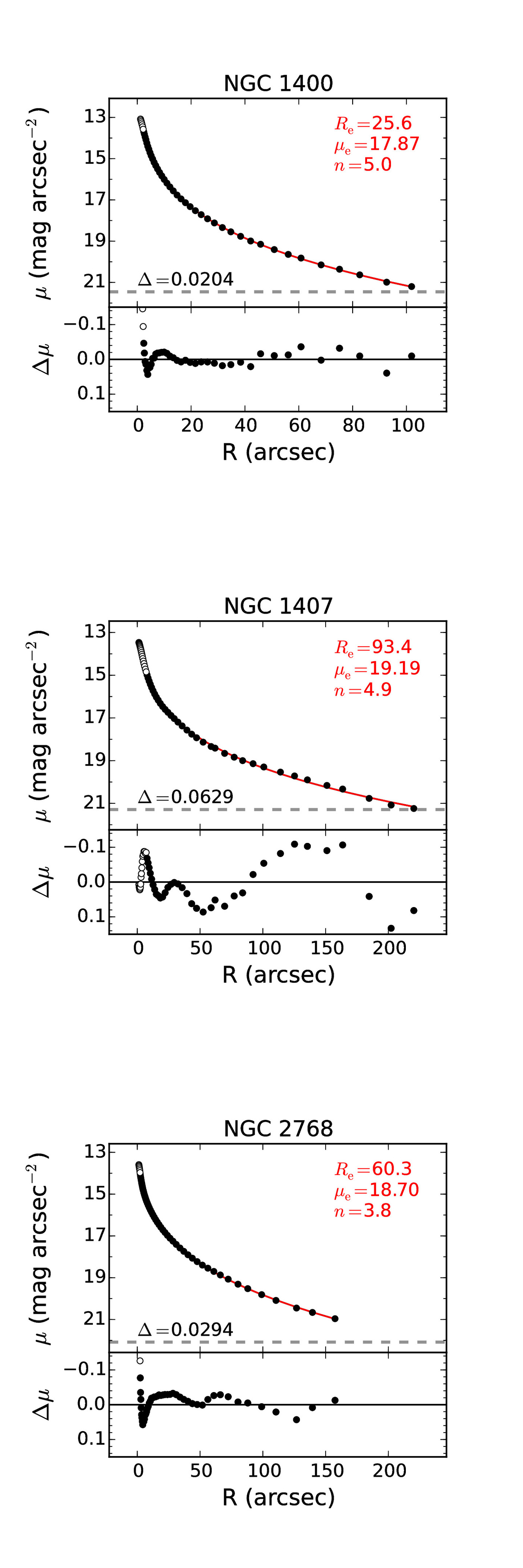}
	    \caption{\label{fig:corrGC2} Sersic fit to 3.6$\mu$m surface brightness profile and residuals as a function of circular equivalent radius. The {\it upper} panel shows the data points (with  
	    excluded data points shown by open circles) and the best fit Sersic profile in red. Parameters for the Sersic fit are given in the top right.  The dashed line 
	    shows 3$\times$ the rms of the sky background level. $\Delta$ gives the rms of the residuals in mag arcsec$^{-2}$. The {\it lower} panel shows the residuals of the Sersic model fit minus the surface brightness data. 
}

\end{figure} 

\begin{figure}
        \includegraphics[angle=0, width=0.40\textwidth]{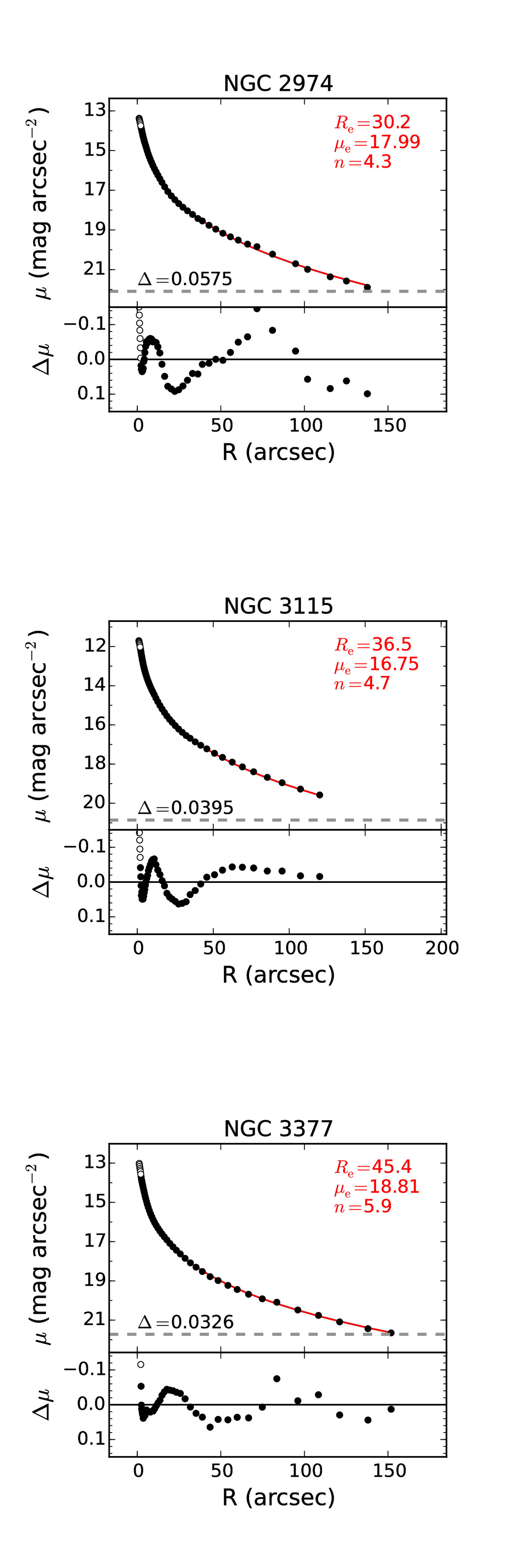}
	    \caption{\label{fig:corrGC2} Sersic fit to 3.6$\mu$m surface brightness profile and residuals as a function of circular equivalent radius. The {\it upper} panel shows the data points (with  
	    excluded data points shown by open circles) and the best fit Sersic profile in red. Parameters for the Sersic fit are given in the top right.  The dashed line 
	    shows 3$\times$ the rms of the sky background level. $\Delta$ gives the rms of the residuals in mag arcsec$^{-2}$. The {\it lower} panel shows the residuals of the Sersic model fit minus the surface brightness data. 
}

\end{figure}

\begin{figure}
        \includegraphics[angle=0, width=0.40\textwidth]{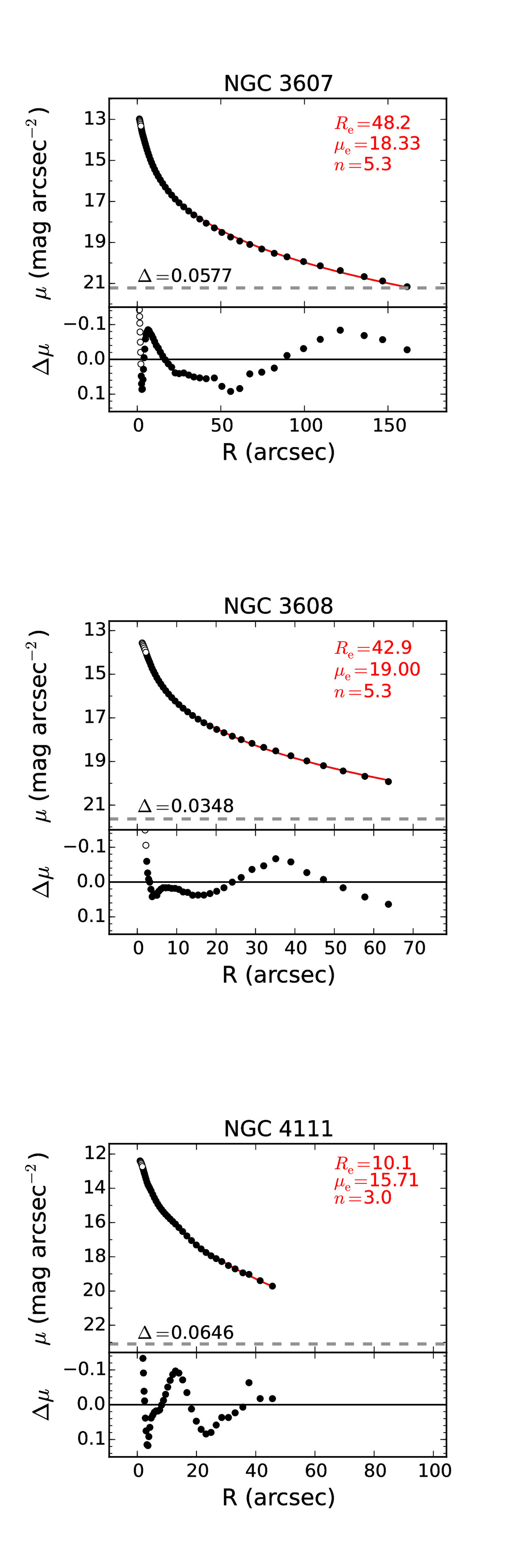}
	    \caption{\label{fig:corrGC2} Sersic fit to 3.6$\mu$m surface brightness profile and residuals as a function of circular equivalent radius. The {\it upper} panel shows the data points (with  
	    excluded data points shown by open circles) and the best fit Sersic profile in red. Parameters for the Sersic fit are given in the top right.  The dashed line 
	    shows 3$\times$ the rms of the sky background level. $\Delta$ gives the rms of the residuals in mag arcsec$^{-2}$. The {\it lower} panel shows the residuals of the Sersic model fit minus the surface brightness data. 
}

\end{figure}

\begin{figure}
        \includegraphics[angle=0, width=0.40\textwidth]{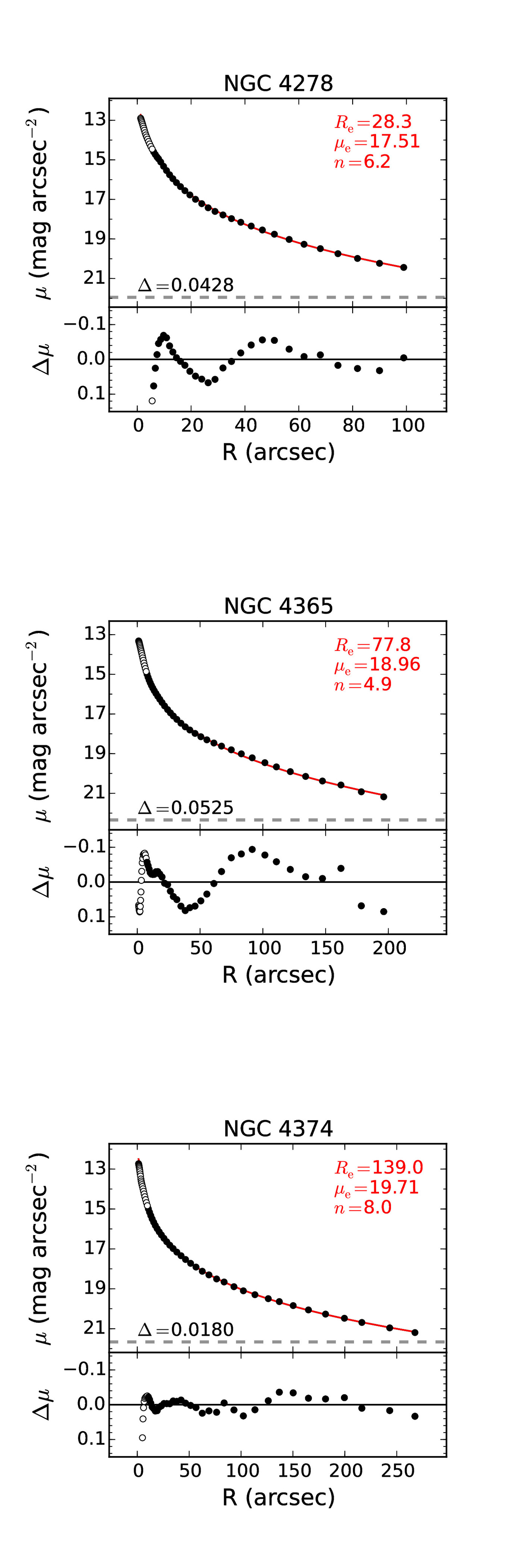}
	    \caption{\label{fig:corrGC2} Sersic fit to 3.6$\mu$m surface brightness profile and residuals as a function of circular equivalent radius. The {\it upper} panel shows the data points (with  
	    excluded data points shown by open circles) and the best fit Sersic profile in red. Parameters for the Sersic fit are given in the top right.  The dashed line 
	    shows 3$\times$ the rms of the sky background level. $\Delta$ gives the rms of the residuals in mag arcsec$^{-2}$. The {\it lower} panel shows the residuals of the Sersic model fit minus the surface brightness data. 
}

\end{figure} 

\begin{figure}
        \includegraphics[angle=0, width=0.40\textwidth]{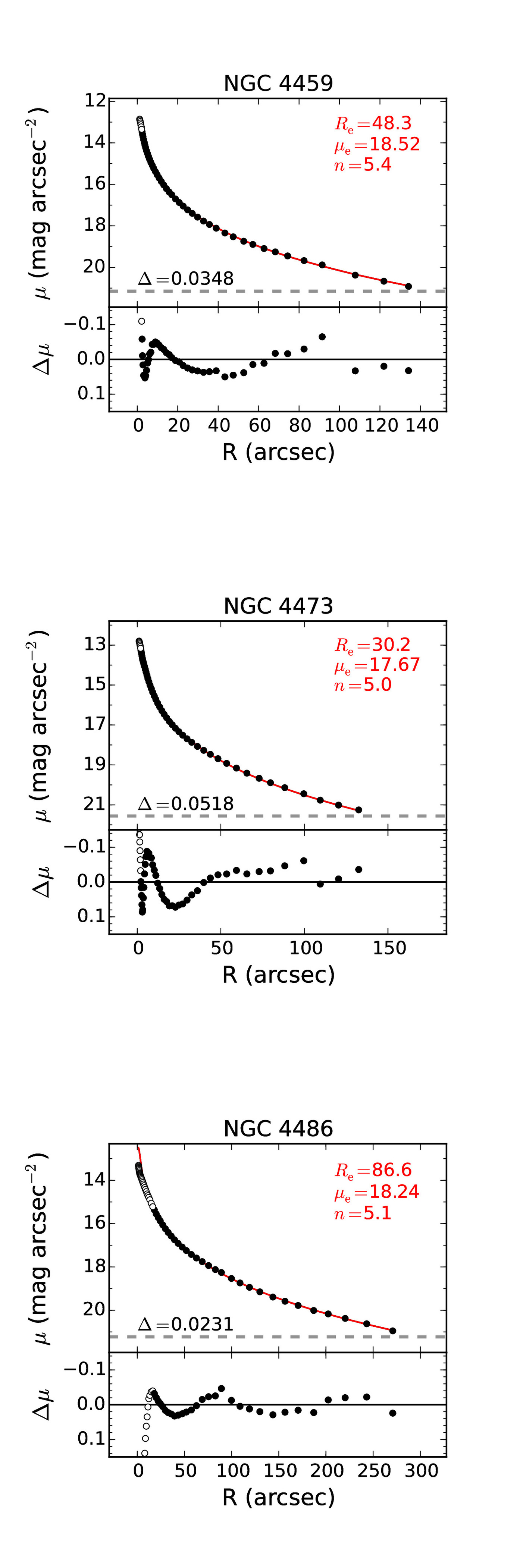}
	    \caption{\label{fig:corrGC2} Sersic fit to 3.6$\mu$m surface brightness profile and residuals as a function of circular equivalent radius. The {\it upper} panel shows the data points (with  
	    excluded data points shown by open circles) and the best fit Sersic profile in red. Parameters for the Sersic fit are given in the top right.  The dashed line 
	    shows 3$\times$ the rms of the sky background level. $\Delta$ gives the rms of the residuals in mag arcsec$^{-2}$. The {\it lower} panel shows the residuals of the Sersic model fit minus the surface brightness data. 
}

\end{figure}

\begin{figure}
        \includegraphics[angle=0, width=0.40\textwidth]{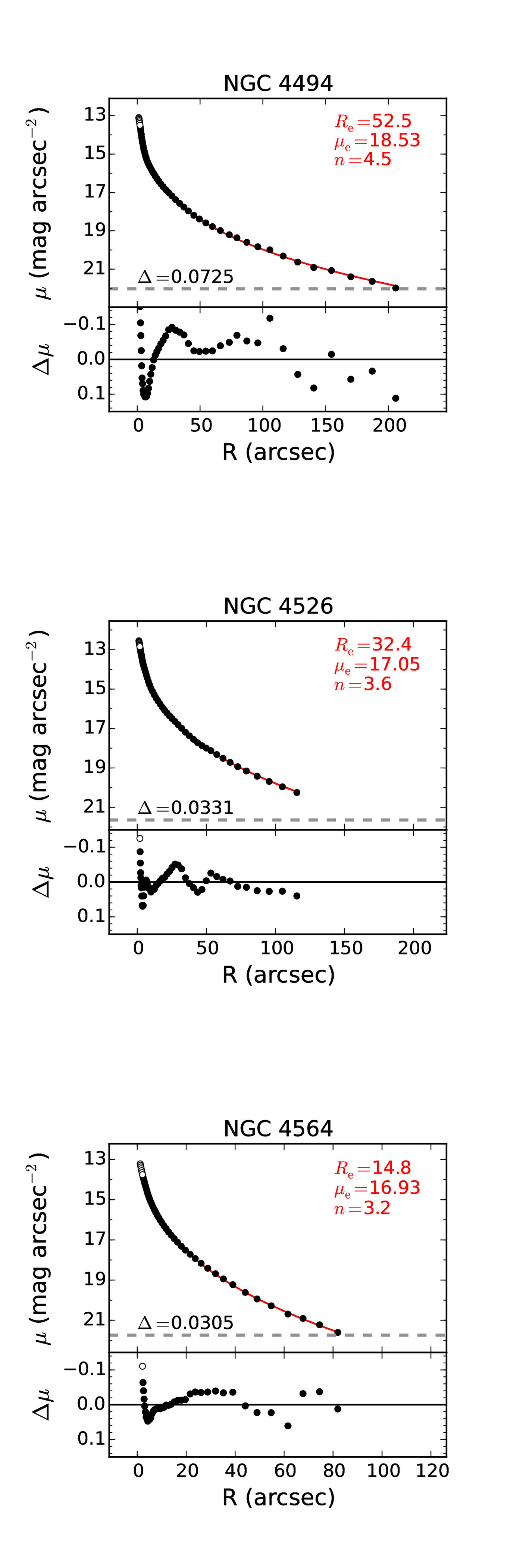}
	    \caption{\label{fig:corrGC2} Sersic fit to 3.6$\mu$m surface brightness profile and residuals as a function of circular equivalent radius. The {\it upper} panel shows the data points (with  
	    excluded data points shown by open circles) and the best fit Sersic profile in red. Parameters for the Sersic fit are given in the top right.  The dashed line 
	    shows 3$\times$ the rms of the sky background level. $\Delta$ gives the rms of the residuals in mag arcsec$^{-2}$. The {\it lower} panel shows the residuals of the Sersic model fit minus the surface brightness data. 
}

\end{figure} 

\begin{figure}
        \includegraphics[angle=0, width=0.40\textwidth]{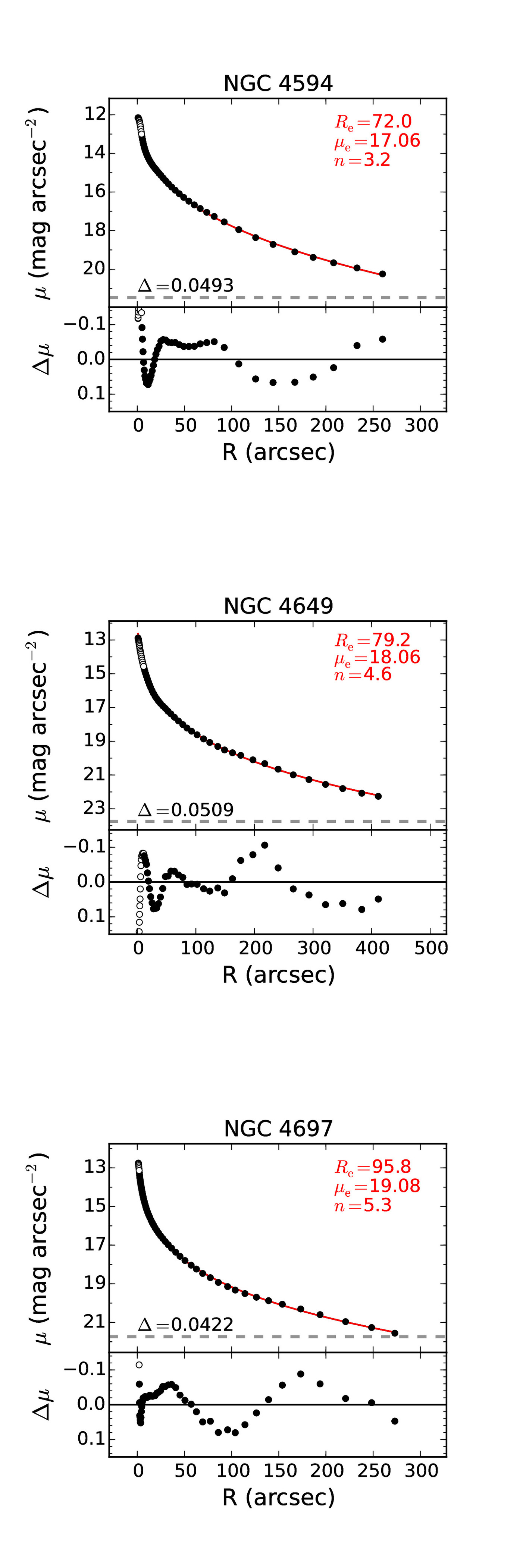}
	    \caption{\label{fig:corrGC2} Sersic fit to 3.6$\mu$m surface brightness profile and residuals as a function of circular equivalent radius. The {\it upper} panel shows the data points (with  
	    excluded data points shown by open circles) and the best fit Sersic profile in red. Parameters for the Sersic fit are given in the top right.  The dashed line 
	    shows 3$\times$ the rms of the sky background level. $\Delta$ gives the rms of the residuals in mag arcsec$^{-2}$. The {\it lower} panel shows the residuals of the Sersic model fit minus the surface brightness data. 
}

\end{figure}

\begin{figure}
        \includegraphics[angle=0, width=0.40\textwidth]{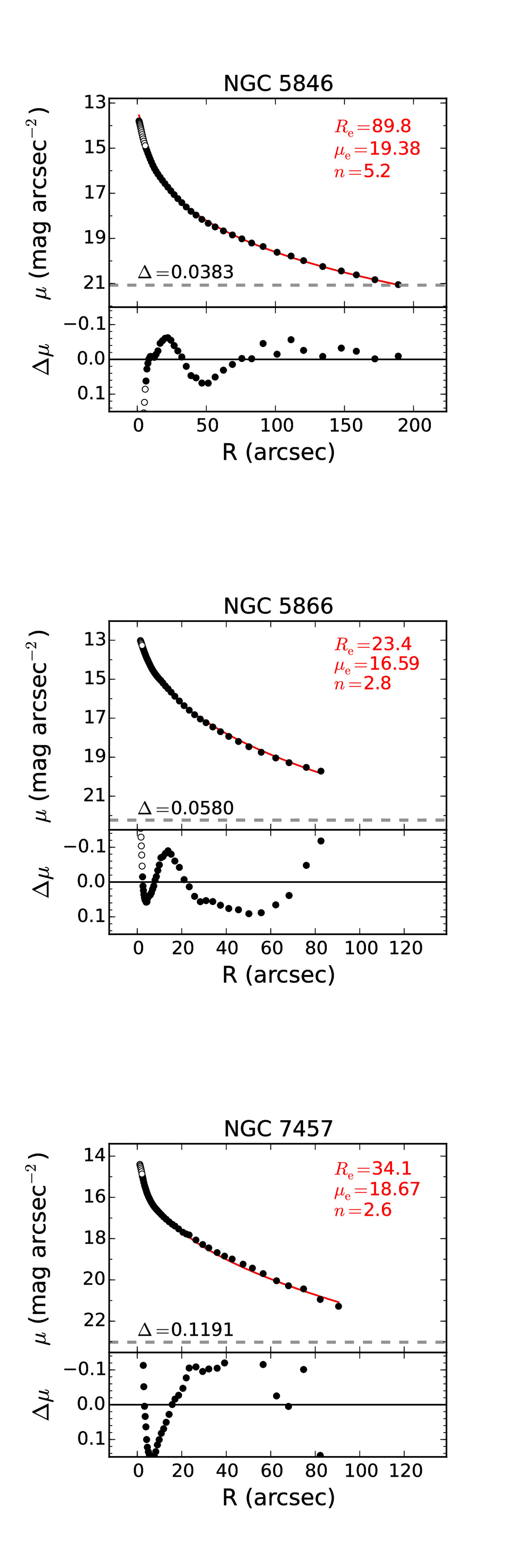}
	    \caption{\label{fig:corrGC2} Sersic fit to 3.6$\mu$m surface brightness profile and residuals as a function of circular equivalent radius. The {\it upper} panel shows the data points (with  
	    excluded data points shown by open circles) and the best fit Sersic profile in red. Parameters for the Sersic fit are given in the top right.  The dashed line 
	    shows 3$\times$ the rms of the sky background level. $\Delta$ gives the rms of the residuals in mag arcsec$^{-2}$. The {\it lower} panel shows the residuals of the Sersic model fit minus the surface brightness data. 
}

\end{figure}

\section{Appendix C}

Six additional (non-SLUGGS) galaxies have been measured in this study using the 
procedure described above. In Table 4 we list the Astronomical Observation Requests 
and in Table 5 the measured 3.6$\mu$m properties for these additional galaxies. Figures 15 and 16 show 
Sersic fits to their 3.6$\mu$m surface brightness profiles.

\begin{table*}
\centering
{\small \caption{Spitzer Space Telescope Astronomical Observation Requests for non-SLUGGS Galaxies}}
\begin{tabular}{@{}cl}
\hline
\hline
Galaxy & Astronomical Observation Requests\\
$\rm [NGC]$ & \\
\hline

1052 &
	r11516672,
	r49600512,
	r49612288\\

2549	 &
	r26602240,
	r49447424,
	r49447680,
	r49619712\\
	
3379 &
	r4445696,
	r49411584,
	r49411840,
	r50629376,
	r50629632,
	r50629888,
	r52832512,
	r52832768,
	r52833024,
	r52833280,
	r52833536,
	r52833792,
	r52834048\\

3665 &
	r49465344,
	r49465600\\
	
3998 &
	r4452608,
	r42242560,
	r42242816,
	r49622784,
	r50586368,
	r50586624,
	r50586880,
	r52892416,
	r52892672,
	r52892928,
	r52893184,
	r52893440,
	r52893696\\

4551 &
	r49510400,
	r49510656\\
		
\hline
\end{tabular}
\begin{flushleft}
{\small 
}

\end{flushleft}
\end{table*}

\begin{table*}
\centering
{\small \caption{Non-SLUGGS galaxy properties}}
\begin{tabular}{@{}cccccccccc}
\hline
\hline
Galaxy & Type & Core & Dist. & Age & m$_{3.6}$ & M$_{\ast}$ & R$_e$ & $\mu_e$ & n\\
$\rm [NGC]$ & & & [Mpc] &[Gyr] &  [mag]  &  [M$_{\odot}$] & [arcsec]  & [mag/arcsec$^2$] & \\
(1) & (2) & (3) & (4) & (5) & (6) & (7) & (8) & (9) & (10)\\\\
\hline
1052 & E3-4/S0 & 1 & 19.4 & 13.0 & 7.12 & 11.02 & 21.9 & 17.16 & 3.4\\
2549 & S0 & 3 & 12.3 & 8.9 & 7.75  & 10.28 & 14.7 & 16.88 & 3.1\\
3379 & E0-1 & 1 & 10.3 & 13.7 & 5.92  & 10.96 & 54.9 & 18.02 & 5.7\\
3665 & S0 & -- & 33.1 & 13.2 & 7.12  & 11.48 & 50.5 & 18.95 & 5.4\\
3998 & S0 & 2 & 13.7 & 13.7 & 7.04  & 10.76 & 19.1 & 16.92 & 4.0\\
4551 & E & 3 & 16.1 & 13.2 & 8.67  & 10.24 & 13.8 & 17.45 & 2.1\\
\hline
\end{tabular}
\begin{flushleft}
{\small 

Notes: columns are (1) galaxy name, (2) Hubble type, (3)  1= core, 2 = intermediate, 3 = cusp central light profile, (4) distance, (5) mean stellar age from McDermid et al. (2015), except for NGC 1052 from Milone et al. (2007), (6) 3.6 micron apparent mag., (7) stellar mass,  (8) effective radius, (9) $\mu_e$, (10) Sersic n.
}

\end{flushleft}
\end{table*}

\begin{figure}
        \includegraphics[angle=0, width=0.40\textwidth]{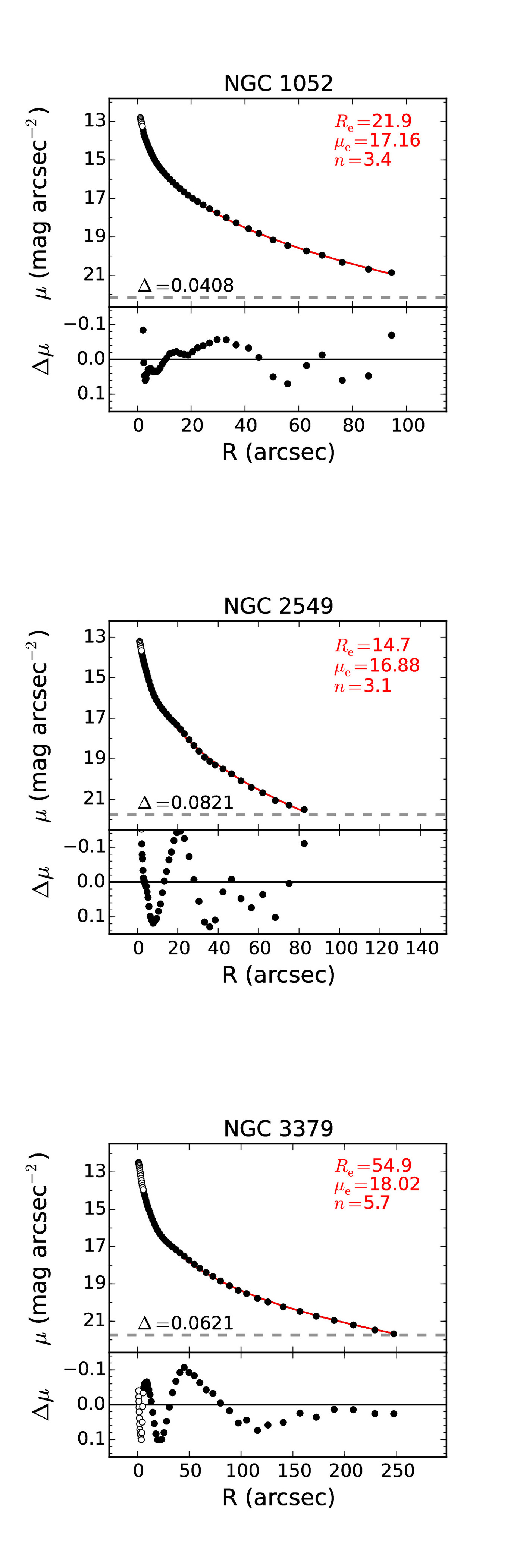}
	    \caption{\label{fig:corrGC2} Sersic fit to 3.6$\mu$m surface brightness profile and residuals as a function of circular equivalent radius for non-SLUGGS galaxies. The {\it upper} panel shows the data points (with  
	    excluded data points shown by open circles) and the best fit Sersic profile in red. Parameters for the Sersic fit are given in the top right.  The dashed line 
	    shows 3$\times$ the rms of the sky background level. $\Delta$ gives the rms of the residuals in mag arcsec$^{-2}$.The {\it lower} panel shows the residuals of the Sersic model fit minus the surface brightness data. 
}
\end{figure}

\begin{figure}
        \includegraphics[angle=0, width=0.40\textwidth]{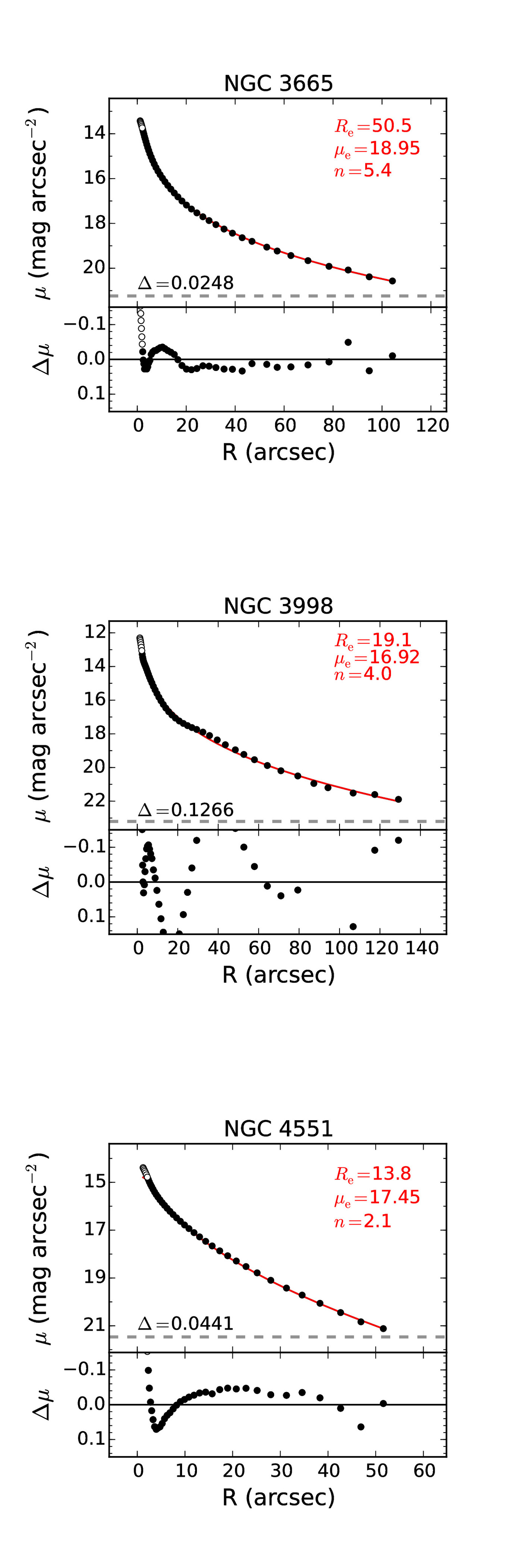}
	    \caption{\label{fig:corrGC2} Sersic fit to 3.6$\mu$m surface brightness profile and residuals as a function of circular equivalent radius for non-SLUGGS galaxies. The {\it upper} panel shows the data points (with  
	    excluded data points shown by open circles) and the best fit Sersic profile in red. Parameters for the Sersic fit are given in the top right.  The dashed line 
	    shows 3$\times$ the rms of the sky background level. $\Delta$ gives the rms of the residuals in mag arcsec$^{-2}$.The {\it lower} panel shows the residuals of the Sersic model fit minus the surface brightness data. 
}
\end{figure}

\end{document}